\documentclass[aps,prl,reprint,balancelastpage,nofootinbib,preprintnumbers,superscriptaddress]{revtex4-1}

\usepackage{slashed}
\usepackage{bm}
\usepackage{latexsym,amssymb,amsmath,float,url,mathrsfs}
\usepackage{latexsym}
\usepackage{graphicx}
\usepackage{epstopdf}
\usepackage{amsmath}
\usepackage{comment}
\usepackage{subfigure}
\usepackage{natbib}
\usepackage{hyperref}
\usepackage{pifont}
\usepackage{blindtext}
\usepackage{adjustbox}
\usepackage{multirow}
\usepackage{tabularx}
\usepackage[table]{xcolor}
\usepackage{orcidlink}
\usepackage{tikz}
\usetikzlibrary{tikzmark}
\usepackage{fontawesome}

\usepackage{blkarray}
\usepackage{graphicx}
\usepackage{amsfonts}
\usepackage{soul}
\usepackage{amssymb}
\usepackage{amsmath}
\usepackage{cancel}
\usepackage{tcolorbox}
\usepackage{tikz-feynman}
\usepackage{tcolorbox}

\usepackage{graphics,appendix,afterpage,makecell} 

\def\dd{{\rm d}}

\definecolor{oucrimsonred}{rgb}{0.6, 0.0, 0.0}
\definecolor{persianblue}{rgb}{0.11, 0.22, 0.73}
\definecolor{forestgreen}{rgb}{0.13,0.35,0.13}
\definecolor{lightgray}{rgb}{0.83, 0.83, 0.83}
 \hypersetup{colorlinks, citecolor=oucrimsonred, linkcolor=black, urlcolor=oucrimsonred}
\definecolor{cornellred}{rgb}{0.7, 0.11, 0.11}
\definecolor{navyblue}{rgb}{0.0, 0.0, 0.5}
\definecolor{amethyst}{rgb}{0.6, 0.4, 0.8}
\definecolor{yellow}{rgb}{1.0, 1.0, 0.0}
\definecolor{firebrick}{rgb}{0.7, 0.13, 0.13}
\definecolor{tangerineyellow}{rgb}{1.0, 0.8, 0.0}
\definecolor{deepfuchsia}{rgb}{0.76, 0.33, 0.76}
\definecolor{amber}{rgb}{1.0, 0.75, 0.0}
\definecolor{VioletRed4}{rgb}{0.55, 0.13, .32}
\definecolor{indiagreen}{rgb}{0.07, 0.53, 0.03}
\definecolor{VioletRed4}{rgb}{0.55, 0.13, .32}
\newcommand{\be}{\begin{equation}}
\newcommand{\ee}{\end{equation}}
\newcommand{\bea}{\begin{equation} \begin{aligned}}
\newcommand{\eea}{\end{aligned} \end{equation}}

\definecolor{oucrimsonred}{rgb}{0.6, 0.0, 0.0}
\newcommand\vertarrowbox[3][6ex]{%
  \begin{array}[t]{@{}c@{}} #2 \\
  \left\uparrow\vcenter{\hrule height #1}\right.\kern-\nulldelimiterspace\\
  \makebox[0pt]{\scriptsize#3}
  \end{array}%
}

\def\g{{\text{\tiny g}}}

\newcommand{\Pz}{\mathcal P_{\zeta_{\text{\tiny g}}}}

\newcommand{\OGW}{\Omega_\text{GW}}

\definecolor{verdechiaro}{rgb}{0.6,1,0.6}
\definecolor{giallochiaro}{rgb}{1,1,0.6}
\definecolor{bluscuro}{rgb}{0.15, 0.2, 0.9}
\definecolor{verdes}{rgb}{0.1, 0.5, 0.1}%
\definecolor{tangerineyellow}{rgb}{1.0, 0.8, 0.0}

\definecolor{americanrose}{rgb}{1.0, 0.01, 0.24}
\definecolor{cobalt}{rgb}{0.0, 0.28, 0.67}
\definecolor{brandeisblue}{rgb}{0.0, 0.44, 1.0}
\definecolor{mycolor}{rgb}{0.0, 0.0, 0.5}
\definecolor{oxfordblue}{rgb}{0.0, 0.13, 0.28}
\definecolor{azure}{rgb}{0.0, 0.5, 1.0}
\definecolor{turquoiseblue}{rgb}{0.0, 1.0, 0.94}
\newtcolorbox{mynewbox}[1]{colback=white!5!white,colframe=azure!75!black,fonttitle=\bfseries,title=#1}
\newtcolorbox{mybox}{colback=mycolor!5!white,colframe=azure!75!black}
\newtcolorbox{mynamedbox}[1]{colback=mycolor!5!white,colframe=azure!75!black,title=#1}
\definecolor{venetianred}{rgb}{0.78, 0.03, 0.08}
\newtcolorbox{mynamedbox1}[1]{colback=venetianred!5!white,colframe=venetianred!80!black,title=#1}
\newtcolorbox{mynamedbox2}[1]{colback=azure!5!white,colframe=azure!80!black,title=#1}

\definecolor{verdes}{rgb}{0.1, 0.5, 0.1}%
\definecolor{cornellred}{rgb}{0.7, 0.11, 0.11}

\newcommand{\nn}{\nonumber}

\definecolor{VioletRed4}{rgb}{0.55, 0.13, .32}

\hypersetup{
     colorlinks   = true,
     citecolor    = violet,
     urlcolor     = violet,
     linkcolor    = violet}

\definecolor{rossocorsa}{rgb}{0.83, 0.0, 0.0}

\def\lsim{\mathrel{\rlap{\lower4pt\hbox{\hskip0.5pt$\sim$}}
    \raise1pt\hbox{$<$}}}         
\def\gsim{\mathrel{\rlap{\lower4pt\hbox{\hskip0.5pt$\sim$}}
    \raise1pt\hbox{$>$}}}         

\usepackage[normalem]{ulem}

\begin{document}

\title[]{How Well Do We Know the Scalar-Induced Gravitational Waves?}

\author{A. J. Iovino\orcidlink{0000-0002-8531-5962}}
\affiliation{Dipartimento di Fisica, ``Sapienza'' Universit\`a di Roma, Piazzale Aldo Moro 5, 00185, Roma, Italy}
\affiliation{INFN Sezione di Roma, Piazzale Aldo Moro 5, 00185, Roma, Italy}
\affiliation{Department of Theoretical Physics and Gravitational Wave Science Center,  \\
24 quai E. Ansermet, CH-1211 Geneva 4, Switzerland}

\author{S. Matarrese\orcidlink{0000-0002-2573-1243}}
\affiliation{Dipartimento di Fisica e Astronomia ``Galileo Galilei'', Universit\`a degli Studi di Padova, Via Marzolo 8, I-35131, Padova, Italy}
\affiliation{INFN, Sezione di Padova, Via Marzolo 8, I-35131, Padova, Italy}
\affiliation{INAF - Osservatorio Astronomico di Padova, Vicolo dell'Osservatorio 5, I-35122 Padova, Italy}
\affiliation{Gran Sasso Science Institute, Viale F. Crispi 7, I-67100 L'Aquila, Italy}

\author{G. Perna\orcidlink{0000-0002-7364-1904}}
\affiliation{Dipartimento di Fisica e Astronomia ``Galileo Galilei'', Universit\`a degli Studi di Padova, Via Marzolo 8, I-35131, Padova, Italy}
\affiliation{INFN, Sezione di Padova, Via Marzolo 8, I-35131, Padova, Italy}
\affiliation{Department of Theoretical Physics and Gravitational Wave Science Center,  \\
24 quai E. Ansermet, CH-1211 Geneva 4, Switzerland}

\author{A. Ricciardone\orcidlink{0000-0002-5688-455X}}
\affiliation{Dipartimento di Fisica ``Enrico Fermi'', Universit\`a di Pisa, Largo Bruno Pontecorvo 3, Pisa I-56127, Italy}
\affiliation{INFN, Sezione di Pisa, Largo Bruno Pontecorvo 3, Pisa I-56127, Italy}

\author{A. Riotto\orcidlink{0000-0001-6948-0856}}
\affiliation{Department of Theoretical Physics and Gravitational Wave Science Center,  \\
24 quai E. Ansermet, CH-1211 Geneva 4, Switzerland}


\begin{abstract}
\noindent
Gravitational waves sourced by amplified scalar perturbations are a common prediction across a wide range of cosmological models. These scalar curvature fluctuations are inherently nonlinear and typically non-Gaussian. We argue that the effects of non-Gaussianity may not always be adequately captured by an expansion around a Gaussian field, expressed through nonlinear parameters such as $f_{\text{\tiny NL}}$. As a consequence, the resulting amplitude of the stochastic gravitational wave background may differ significantly from predictions based on the standard quadratic source model routinely used in the literature. 

\end{abstract}

\maketitle

\hspace{-0.35cm}\noindent\textbf{Introduction --} 
One of the most challenging goals of current and future gravitational wave observatories is to investigate the presence of scalar-induced gravitational waves (SIGWs). This stochastic gravitational wave background, which is a standard prediction of General Relativity, arises from  scalar fluctuations which are amplified compared to the initial seeds giving rise to the large-scale structure~\cite{Tomita:1975kj, Matarrese:1993zf, Acquaviva:2002ud, Mollerach:2003nq, Ananda:2006af, Baumann:2007zm}. In order to have a sizable effect, the scalar fluctuations must be enhanced with respect to those observed at cosmic microwave background  scales. Many theoretical frameworks of the early universe predict an enhancement of curvature perturbations on small scales, leading to the generation of SIGWs with distinctive signatures. These predictions stem from scenarios such as single-field inflation with ultra-slow-roll phases~\cite{Ivanov:1994pa,Leach:2000ea,Bugaev:2008gw,Alabidi:2009bk,Drees:2011hb,Drees:2011yz,Alabidi:2012ex,
Ballesteros:2017fsr,Di:2017ndc,Germani:2017bcs,Cicoli:2018asa,Ozsoy:2018flq,Bhaumik:2019tvl,Ballesteros:2020qam,Karam:2022nym,Franciolini:2022pav},
multifield dynamics~\cite{Garcia-Bellido:1996mdl,Bugaev:2011wy,Kawasaki:2015ppx,Clesse:2015wea,Braglia:2020eai,Palma:2020ejf,Braglia:2022phb}
and curvaton models~\cite{Kawasaki:2012wr,
Ando:2017veq,Ando:2018nge,Ferrante:2022mui,Ferrante:2023bgz,Kawasaki:2021ycf,Chen:2019zza,Liu:2020zzv,Inomata:2020xad,Pi:2021dft, Gow:2023zzp,Inomata:2023drn}. 
The detection and characterization of SIGWs would provide direct insight into the processes shaping the primordial universe. Moreover, they are strictly linked to the formation of primordial black holes (for a review, see Ref. \cite{LISACosmologyWorkingGroup:2023njw}).

The curvature perturbations $\zeta$ giving rise to SIGWs are typically nonlinear and exhibit non-Gaussian characteristics\,\cite{Bartolo:2004if,Sasaki:2006kq,Cai:2018dkf,Atal:2019cdz,Figueroa:2020jkf,Biagetti:2021eep,Pi:2022ysn,Cai:2022erk,Gow:2022jfb,Hooshangi:2023kss,Ballesteros:2024pwn}.
In the literature it is common to take a phenomenological approach and expand the curvature perturbation around its Gaussian counterpart \cite{Gangui:1993tt,Komatsu:2001rj}
\be
\label{fnl}
\zeta = \zeta_{\text{\tiny g}} + \frac{3}{5} f_{\text{\tiny NL}} \zeta_{\text{\tiny g}}^2 + \cdots,
\ee
where $f_{\text{\tiny NL}}$ (and higher-order terms) is supposed to parametrize the amount of non-Gaussianity. The resulting effects on the SIGW spectrum have been widely studied in literature~\cite{Cai:2018dkf, Cai:2018dig, Unal:2018yaa, Cai:2019elf,  Hajkarim:2019nbx, Atal:2021jyo, Yuan:2020iwf, Domenech:2021and, Garcia-Saenz:2022tzu, Liu:2023ymk,  Yuan:2023ofl, Li:2023xtl, Adshead:2021hnm, Perna:2024ehx,Inui:2024fgk}. 

In this Letter we argue that the effects of non-Gaussianity on SIGWs may not always be accurately captured by an expansion around a Gaussian field. Consequently, the amplitude of the resulting SGWB could differ significantly from predictions based on the conventional quadratic source framework.

Our conclusions hinge on two arguments. The first aspect concerns the form of the fully nonlinear source on superhorizon scales, which establish the initial condition for the generation of SIGWs as curvature perturbations re-enter the horizon. The second focuses on the estimation of the impact of nonlinearities when the wavelength of the curvature perturbation becomes smaller than the Hubble radius. We address these two points sequentially.   

\vskip 0.5cm
\hspace{-0.35cm}\noindent\textbf{A hint from the fully nonlinear superhorizon source --} 
To gauge the impact of nonlinearities in curvature perturbations on the SIGWs produced during the radiation-dominated era, we start from the long-wavelength limit of Einstein's equations while retaining full nonlinearity in the perturbations. In the comoving uniform-energy-density gauge, the tensor modes $\gamma_{ij}$ are expressed as~\cite{Harada:2015yda}
\bea\label{eq:hij}
&\gamma_{ij}(t,{\bf x})=-\frac{1}{3} P_{ijkl}^{\mathrm{TT}}S_{kl}({\bf x})\left(\frac{1}{a H}\right)^2,
\eea
where \( a \) is the scale-factor, \( H \) the Hubble parameter, and the transverse and traceless projector reads
\begin{eqnarray}
P_{ijkl}^{\mathrm{TT}} &=& \frac{1}{2}\left(P_{ik} P_{jl} + P_{jk} P_{il} - P_{ij} P_{kl}\right), \nonumber\\
P_{ij} &=& \delta_{ij} - \frac{\partial_i \partial_j}{\partial^2}.
\end{eqnarray}
The source term \( S_{ij} \) is given by\,\cite{Harada:2015yda} 
\begin{eqnarray}
\label{Eq::p_ij}
S_{ij} &=& \frac{1}{\psi^4}\left[-\frac{2}{\psi}\left(\partial_i \partial_j \psi - \frac{1}{3}\delta_{ij} \nabla^2\psi\right) \right. \nonumber\\
&+&\left. \frac{6}{\psi^2} \left(\partial_i \psi \partial_j \psi - \frac{1}{3}\delta_{ij}\partial^k \psi \partial_k \psi \right) \right].
\end{eqnarray}
The gravitational potential \( \psi \) relates to the nonlinear curvature perturbation \( \zeta \) through\,\cite{Lyth:2004gb,Harada:2015yda}
\be
\psi({\bf x}) = e^{\zeta({\bf x})/2}.
\ee
By introducing a variable transformation, 
\be
Y({\bf x}) = \frac{1}{\psi^2({\bf x})}, 
\ee
the source term can be elegantly rewritten as 
\begin{equation}
\label{Eq::Source}
S_{ij} = -\partial_i Y \partial_j Y+\partial_i(Y\partial_j Y) -\frac{1}{3} \delta_{ij} Y\nabla^2Y.
\end{equation}
Notably,  only the first term survives under the action of the projector
$P_{ijkl}^{\mathrm{TT}}$ and we will disregard the others from now on.

We assume that the constant (in time) nonlinear curvature perturbation \( \zeta \) on superhorizon scales is connected to its Gaussian counterpart \( \zeta_{\text{\tiny g}} \) via the relation (where the absolute value has a nontrivial origin, as discussed in the Supplementary Material)
\be
\label{eq:NG}
\zeta = -\mu \ln\left|1-\frac{\zeta_{\text{\tiny g}}}{\mu}\right|\,.
\ee
This relation arises naturally in standard models such as the ultra-slow-roll inflationary scenario~\cite{Atal:2019cdz} and the curvaton mechanism~\cite{Sasaki:2006kq,Pi:2022ysn}. In ultra-slow-roll scenarios, \( \mu \) is typically related to the slope of the inflationary power spectrum of \( \zeta_{\text{\tiny g}} \) beyond its peak, with values ranging between \( 1.5 \) and \( 6.6 \)~\cite{Atal:2018neu,Karam:2022nym,Frosina:2023nxu}. For curvaton models, \( \mu \) is intricately determined by the time of the curvaton's decay into radiation and the shape of the power spectrum~\cite{Lyth:2002my,Enqvist:2005pg,Lyth:2006gd,Ando:2018nge,Ferrante:2023bgz}, with typical values of \( \mathcal{O}(0.1) \).
Expanding Eq.~\eqref{eq:NG} to second order and comparing it with the common series expansion \eqref{fnl}, we find that \( f_{\text{\tiny NL}} = 5/(6\mu) \). To be model independent, we will treat \( \mu \) as a free parameter from now on.

In terms of \( \zeta_{\text{\tiny g}} \), the relevant term in the source  \eqref{Eq::Source}  which is not projected away is  
\be
\label{generalzeta}
S_{ij} = -\left|1-\frac{\zeta_{\text{\tiny g}}}{\mu}\right|^{2(\mu-1)}\partial_i \zeta_{\text{\tiny g}} \partial_j \zeta_{\text{\tiny g}}\,.
\ee
For large \( \mu \) (small non-Gaussianity), we recover the standard quadratic source
\be
\lim_{\mu \to \infty} \left|1-\frac{\zeta_{\text{\tiny g}}}{\mu}\right|^{2(\mu-1)} = e^{-2\zeta_{\text{\tiny g}}} \simeq 1\,,
\ee
where in the last passage we have taken $\zeta_{\text{\tiny g}}\ll 1$.
Conversely, for small \( \mu \) (corresponding to large non-Gaussianity), the source becomes
\be
\label{generalzetaNG}
S_{ij} = -\left|1-\frac{\zeta_{\text{\tiny g}}}{\mu}\right|^{-2}\partial_i \zeta_{\text{\tiny g}} \partial_j \zeta_{\text{\tiny g}}\,.
\ee
This prefactor highlights the possibility of significant deviations from the standard quadratic source while remaining within the perturbative regime: significant enhancements occur if \( \zeta_{\text{\tiny g}} \sim \mu \ll 1 \), whereas large suppressions for \( 1 \gg \zeta_{\text{\tiny g}} \gg \mu \).

Notably, the case with \( \mu = 1 \) precisely reproduces the superhorizon quadratic source made of Gaussian fields, even though it corresponds to \( f_{\text{\tiny NL}} = 5/6 \), which represents a relatively sizable non-Gaussian perturbation. 

These findings show that the SIGW source, just before the perturbations re-enter the horizon, may deviate significantly from predictions based on Gaussian fields. Moreover, the  non-Gaussian effects may not be captured by an expansion in  \( f_{\text{\tiny NL}} \) or higher-order nonlinear parameters.

\vskip 0.5cm
\hspace{-0.5cm} \textbf{The impact of nonlinearities on the SIGW: an estimate --} 
A comprehensive and analytical calculation of the impact of curvature perturbation nonlinearities on SIGWs,  including the subhorizon evolution, is an Herculean task.  Admittedly,  we have not been able to achieve it. The difficulty of the problem  lies in several aspects: the impossibility of calculating the Fourier transform of the nonlinear curvature perturbation \eqref{eq:NG}, the ignorance about the nonlinear radiation transfer function to apply when the perturbation are subject to the radiation pressure on sub-horizon scales, to mention a few. 

We should not concede defeat too soon, though. If  we are satisfied with an estimate, we can adopt the line of reasoning based on the following assumptions. First,  we  suppose that the typical values of the nonlinear perturbation $\zeta$ are also in the perturbative regime once the  typical values of the Gaussian field $\zeta_{\text{\tiny g}}$ are in the linear regime, $|\zeta_{\text{\tiny g}}|\ll 1$. For instance, as shown in Fig.\,\ref{fig:zeta}, for $\mu=10^{-1}$ (equivalent to  $f_{\text{\tiny NL}} \simeq 8$) only in a very narrow range of values for $(0.99\lsim \zeta_{\text{\tiny g}}/\mu\lsim 1.01)$, the linearity, i.e. $|\zeta|\ll1$, is violated.

If so, the curvature perturbation which re-enters the horizon may be considered a linear perturbation, its amplitude may be  after all still smaller than unity, even though it is generated by a nonlinear dynamics. The situation here is analogous to what happens to the SIGWs themselves: even though they are generated by a nonlinear source, when they propagate to a detector their amplitude is tiny once they  decouple from the source. They can be therefore considered linear
observables and fully gauge-invariant \cite{DeLuca:2019ufz,Inomata:2019yww,Yuan:2019fwv}.

With this logic in mind, we can estimate the SIGW background by adopting the standard equation for the GWs with a quadratic source in  the Newtonian gauge (the superscript ${}^{\text{\tiny Q}}$  indicating the  source is quadratic, i.e. constructed from the square of linear, but not necessarily Gaussian quantities)
\begin{equation}
\ddot\gamma_{ij}+3 H \dot \gamma_{ij}-\nabla^2 \gamma_{ij}=P_{ijkl}^{\mathrm{TT}} S^{\text{\tiny Q}}_{kl},
\label{eq: eom GW1}
\end{equation}
where dots indicate the derivative with respect to the  cosmic time and in the radiation phase the source is given by~\cite{Acquaviva:2002ud}
\begin{equation}
\label{psi}
S_{ij}^{\text{\tiny Q}}=4\Phi\partial_i\partial_j\Phi+2\partial_i\Phi\partial_j\Phi-\partial_i\left(\frac{\dot \Phi}{ H}+\Phi\right)\partial_j\left(\frac{\dot \Phi}{ H}+\Phi\right)
\end{equation}
in terms of the Bardeen potential $\Phi$. 
Given our assumption of the linearity of the curvature perturbation $\zeta$, the scalar perturbation $\Phi(t,{\bf k})$ in the Newtonian gauge  can be written in terms of the initial gauge-invariant comoving curvature perturbation as
\begin{equation}
\Phi(t,{\bf k}) \equiv \frac 23 T(k,t) \zeta({\bf k}),
\label{eq: Psi to zeta}
\end{equation}
where $T(k,t)$ is the linear transfer function during the radiation domination phase and $\zeta({\bf k})$ is the Fourier transform of the superhorizon curvature perturbation \eqref{eq:NG} which plays the role of the initial condition. 
The technical difficulty we have mentioned earlier arises here, since we are not able to calculate the Fourier transform of the curvature perturbation \eqref{eq:NG} in terms of the one of the Gaussian field $\zeta_{\text{\tiny g}}({\bf k})$. In order to circumvent this problem we make the second assumption. Since 
\be
\partial_i\zeta=\frac{1}{\left|1-\frac{\zeta_{\text{\tiny g}}}{\mu}\right|}\partial_i\zeta_{\text{\tiny g}},
\ee
we approximate the prefactor in front of the gradient term with its 
typical value
\bea
\label{a}
\left|1-\frac{\zeta_{\text{\tiny g}}}{\mu}\right|^{-1}&\simeq& \left|1-\frac{\pm\langle\zeta^2_{\text{\tiny g}}\rangle^{1/2}}{\mu}\right|^{-1}\equiv \mathcal{A},
\eea
where
\be
\langle\zeta^2_{\text{\tiny g}}\rangle=\int\frac{{\rm d}k}{k}\Pz(k)
\ee
and $\Pz(k)$ is the power spectrum of the Gaussian perturbation. In Fig. \ref{fig:zeta} we plot the values of $\zeta$ and ${\cal A}$ as a function of  $\zeta_{\text{\tiny g}}$, where in the case of ${\cal A}$ we intend $\zeta_{\text{\tiny g}}$ to be the typical value $\pm \langle\zeta^2_{\text{\tiny g}}\rangle^{1/2}$. This figure illustrates that $\zeta$ can stay in the linear regime and still have sizable enhancements or suppressions parametrized by the prefactor ${\cal A}$. 
\begin{figure}[t!]
\centering
  \includegraphics[width=0.49\textwidth]{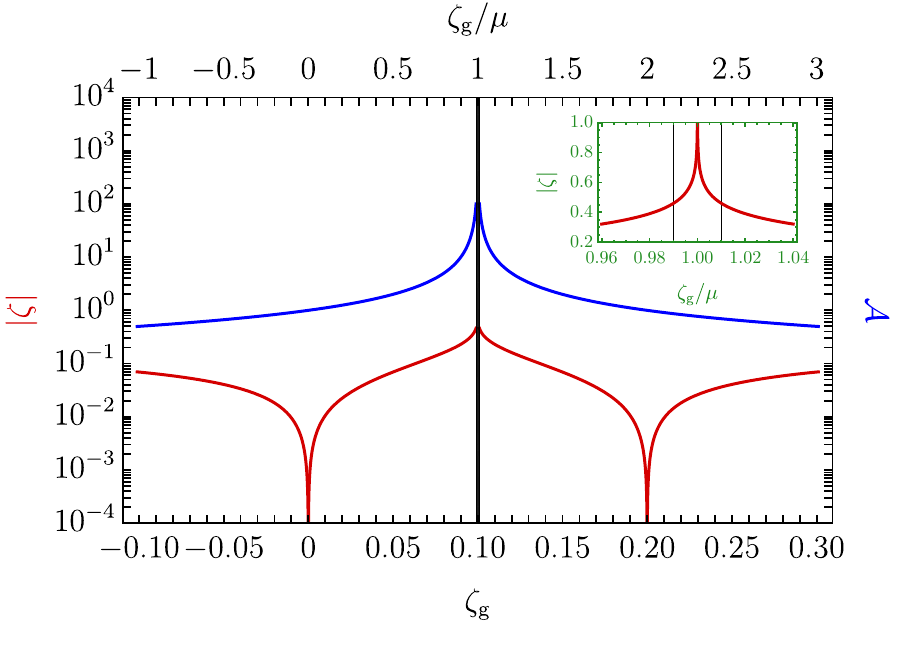}
  \caption{The values of $\zeta$ (red line) and ${\cal A}$ (blue line) as a function of  $\zeta_{\text{\tiny g}}/\mu$, from Eqs. (\ref{eq:NG}) and (\ref{a}), where we set as benchmark value $\mu=0.1$. For the variable ${\cal A}$,  $\zeta_{\text{\tiny g}}$ has to be understood as the typical value $\pm \langle\zeta^2_{\text{\tiny g}}\rangle^{1/2}$. The black stripe shows the values for which linearity, i.e. $|\zeta|\ll 1$, is violated in the case of $\mu=0.1$. The green inset shows the behaviour of $|\zeta|$ in a neighborhood of $\zeta_g/\mu$ where perturbativity is violated.}
  \label{fig:zeta}
\end{figure}
The second approximation is worth the reward  as it allows now to calculate the SIGW background
as
\begin{align}
&\frac{\OGW(f)}{\Omega_{r,0}\,c_g} = \frac{\mathcal{A}^4}{72}
  \int_{-\frac{1}{\sqrt{3}}}^{\frac{1}{\sqrt{3}}}\dd d \int_{\frac{1}{\sqrt{3}}}^{\infty}\hspace{-5pt}\dd s
  \left[ \frac{(d^2-1/3)(s^2-1/3)}{s^2-d^2} \right]^2\nn\\&
\cdot  \Pz\left(\frac{k\sqrt{3}}{2}(s+d)\right) \Pz\left(\frac{k\sqrt{3}}{2}(s-d)\right){\cal I}^2(d,s),
\label{eq: Omega GW with PS0}
\end{align}
where $k=2\pi f$, $\Omega_{r,0}$ parametrises the current density of radiation if the neutrinos were massless, $c_g$ accounts for the change of the effective degrees of freedom of the thermal radiation during the evolution (assuming Standard Model physics), ${\cal I}^2$ is a function derived in Refs.\,\cite{Espinosa:2018eve,Kohri:2018awv} depending on the transfer function. Observe that applying the projector operator to the prefactor would not have affected the overall  behavior, as the operator is dimensionless. 

In Fig.\,\ref{fig:omega} we plot the values of $\OGW(f)$ for different values of $\mu$ and for a lognormal power spectrum 
\be
\Pz(k)=\frac{A}{\sqrt{2\pi\sigma^2}}\,{\rm exp}\left[-\frac{\ln^2(k/k_\star)}{2\sigma^2}\right],
\ee
showing that the SIGW background may differ from the standard calculation based on the quadratic source and the Gaussian curvature perturbation. We have  conservatively excluded those values of $\mu$ which would violate the perturbativity condition on $\zeta$. Of course, our results are not rock-hard as they rely on a series of approximations, but they suggest that the standard calculation of the SIGW background amplitude is not so solid either, hence the question mark in the title. 
\noindent
\begin{figure}[t!]
\centering
\includegraphics[width=0.49\textwidth]{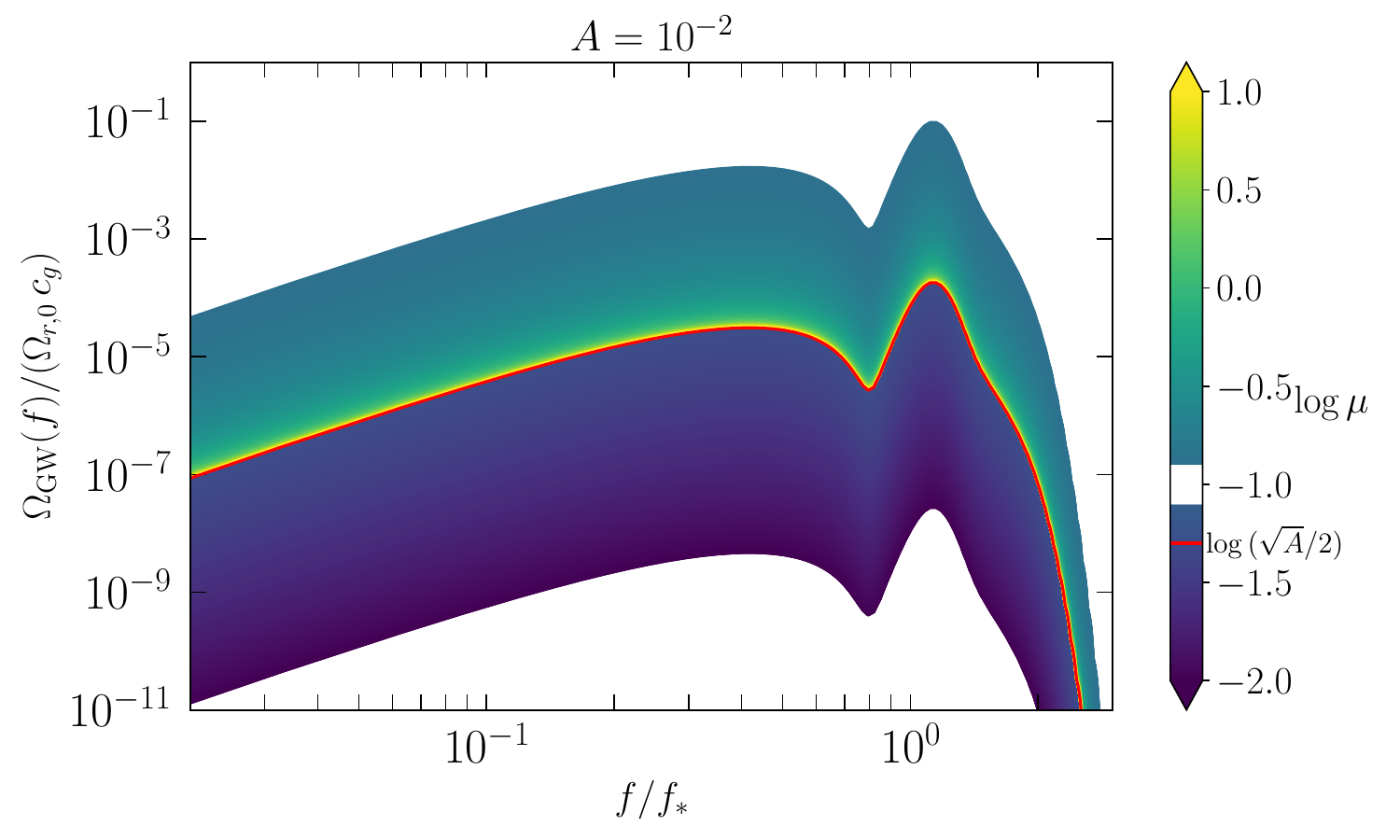}
\caption{Example of SIGW spectrum assuming lognormal template of the scalar power spectrum. We fix the amplitude $A$ to $10^{-2}$, the variance $\sigma$ to $10^{-1}$ and we vary the value of $\mu$. The red line is obtained for $\mathcal{A}=1$, corresponding to the usual standard quadratic spectrum used in the literature  for a radiation dominated universe.}
  \label{fig:omega}
\end{figure}

\vskip 0.5cm
\hspace{-0.55cm} \textbf{Conclusions --} 
The study of SIGWs offers a compelling avenue to probe the nonlinear dynamics of the early universe, providing a potential bridge between enhanced scalar perturbations and phenomena such as primordial black hole formation. By incorporating nonlinear effects into the SIGW source, we uncover potentially significant deviations from predictions rooted in Gaussian curvature perturbations. The behavior of the SIGW background depends intricately on the degree of non-Gaussianity and underscores the potential for both suppression and enhancement relative to standard quadratic models.

Our analysis, though approximate, suggests that even modest nonlinearities can leave observable imprints on the SIGW spectrum, challenging traditional approaches that rely on Gaussian perturbation theory. These findings emphasize the need for more refined models and numerical methods to accurately characterize these effects and their implications for future gravitational wave observatories. As next-generation detectors like LISA\,\cite{LISA:2022kgy} and Einstein Telescope\,\cite{Branchesi:2023mws} promise to improve sensitivity to SIGWs, further exploration of their nonlinear origins will be critical to unlocking new insights into the physics of the early universe. Our findings might have already a significant impact on the interpretation of  the signal seen by pulsar timing arrays  as  originated from a SIGW background and  accompanied by a sizable primordial black hole abundance \cite{Franciolini:2023pbf, Inomata:2023zup, Figueroa:2023zhu,Ellis:2023oxs,You:2023rmn,Balaji:2023ehk,Wang:2023ost,Yi:2023tdk,Domenech:2024rks,Iovino:2024tyg}.

There is of course room to improve  the approach proposed in this Letter. One potential avenue might be to numerically investigate the nonlinear interactions during the ultra-slow roll phase through a lattice simulations. This would allow for the calculation of the source for SIGWs in a fully nonperturbative manner, following techniques similar to those outlined in \cite{Caravano:2024moy}.

\begin{acknowledgments}
\vspace{5pt}\noindent\emph{Acknowledgments --}
We thank V. De Luca, G. Franciolini and D. Racco for useful discussions and comments on the draft.
A.R. acknowledges support from the Swiss National Science Foundation (project number CRSII5\_213497) and from  the Boninchi Foundation for the project ``PBHs in the Era of GW Astronomy''.
\end{acknowledgments}

\bibliography{GWBIB}

\maketitle
\onecolumngrid
\begin{center}
\vspace{0.1in}
 
\vspace{0.05in}
{ \Large\it Supplementary Material}
\end{center}
\onecolumngrid
\setcounter{equation}{0}
\setcounter{figure}{0}
\setcounter{section}{0}
\setcounter{table}{0}
\setcounter{page}{1}
\makeatletter
\renewcommand{\theequation}{S\arabic{equation}}
\renewcommand{\thefigure}{S\arabic{figure}}
\renewcommand{\thetable}{S\arabic{table}}
\section*{How to deal with large curvature fluctuations}
\noindent
The fully non-perturbative expression for the curvature perturbation is on super-horizon scales \cite{Lyth:2004gb} 

\begin{equation}
  \zeta({\bf x},t)=-\Psi({\bf x},t)-\frac{1}{3}\int_{\overline{\rho}}^{\rho({{\bf x},t})}\frac{{\rm d}\rho}{\rho+P},  
\end{equation}
which in  the flat gauge $\Psi=0$ reduces to

\begin{equation}
  \zeta({{\bf x},t})=-\frac{1}{3}\int_{\rho(t)}^{\rho({{\bf x},t})}\frac{{\rm d}\rho}{\rho+P}. 
  \end{equation}
 In the above expressions $\rho$ and $P$ are the energy density and pressure density, respectively, of a fluid and $\Psi$ is the gravitational potential related to the $\psi$ in Eq. (\ref{Eq::p_ij}) by  $\psi=e^{\Psi/2}$, such that in the  the comoving
uniform-energy-density gauge, $\zeta=-\Psi$.

For a canonically normalized scalar field $\rho+P=\dot\phi^2$. In slow-roll one correctly obtains, at the linear level, the standard result

\begin{equation}
\zeta=-\frac{1}{3}\frac{\delta\rho({\bf x},t)}{\dot\phi^2}=-\frac{1}{3}\frac{V'\delta\phi({\bf x},t)}{\dot\phi^2}=-\frac{1}{3}\frac{-3H\dot\phi \,\delta\phi({\bf x},t)}{\dot\phi^2}=H\frac{\delta\phi({\bf x},t)}{\dot\phi}.
 \end{equation}
During a   period of exact ultra-slow roll for which the inflaton potential is totally flat,  one has $\rho+P=\dot\phi^2$. Furthermore $\rho\simeq \dot\phi^2/2$ and therefore (being $t_\star$ denotes the final time the ultra-slow roll stage) \cite{Cai:2018dkf,Biagetti:2018pjj}

\begin{equation}
\label{correct}
  \zeta({{\bf x},t_\star})=-\frac{1}{3}\cdot\frac{1}{2}\int_{\frac{1}{2}\dot{\phi}^2(t_\star)}^{\frac{1}{2}\dot{\phi}^2({{\bf x},t_\star})}\frac{{\rm d}\dot{\phi}^2}{\dot{\phi}^2}=-\frac{1}{6}\ln\frac{\dot{\phi}^2({{\bf x},t_\star})}{\dot{\phi}^2(t_\star)}, 
  \end{equation}
  which is the correct expression. 
 In the literature, however, one   finds the expression

\begin{equation}
  \zeta({{\bf x},t_\star})=-\frac{1}{3}\ln\frac{\dot{\phi}({{\bf x},t_\star})}{\dot{\phi}(t_\star)},
  \end{equation}
which is misleading because, when $\dot{\phi}({{\bf x},t})<0$,  a problem seems to arise when the  argument of the logarithm is negative.
It is also customary to expand the last expression at first order, with $\dot{\phi}({\bf x},t_\star)=\delta\dot{\phi}(t_\star)+\delta\dot{\phi}_\g({\bf x},t_\star)$,

\begin{equation}
  \zeta({{\bf x},t_\star})=-\frac{1}{3}\ln\left(1+\frac{\delta\dot{\phi}_\g({\bf x},t_\star)}{\dot{\phi}(t_\star)}\right),
  \end{equation}
which again indicates a false problem for $\delta\dot{\phi}({\bf x},t_\star)/\dot{\phi}(t_\star)<-1$. One should use instead the expression with an absolute value originated from the correct expression (\ref{correct})

\begin{equation}
  \zeta({{\bf x},t_\star})=-\frac{1}{3}\ln\left|1+\frac{\delta\dot{\phi}_\g({\bf x},t_\star)}{\dot{\phi}(t_\star)}\right|.
  \end{equation}
In ultra-slow roll models one has $\dot{\phi}({\bf x},t_\star)=3H[\phi({\bf x},t)-\phi({\bf x},t_\star)]+\dot{\phi}({\bf x},t)$ \cite{Cai:2018dkf}, and therefore

\begin{equation}
  \zeta({{\bf x},t_\star})=-\frac{1}{3}\ln\left|1-3H\frac{\delta\phi_\g({\bf x},t_\star)}{\dot{\phi}(t_\star)}\right|=-\frac{1}{3}\ln\left|1-3\zeta_\g({\bf x},t_\star)\right|,
\end{equation}
leading to $\mu=1/3$ and the standard result for sharp transitions $f_{\text{\tiny NL}}=5/2$ \cite{Cai:2018dkf}. According to the exact expression (\ref{correct}) there is no obstruction in having values of $\zeta_\g$ larger than $1/3$ and therefore no probability leakage as claimed in Ref. \cite{Escriva:2023uko}. Equivalently, one can safely consider
values of $\zeta_\g$ larger than $\mu$ in Eq. (\ref{eq:NG}).
\end{document}